\documentclass[prl,twocolumn,showpacs,preprintnumbers,amsmath,amssymb]{revtex4}
\usepackage{graphicx}
\usepackage{dcolumn}
\usepackage{bm}

\begin{document}


\title{Galilean Equivalence for Galactic Dark Matter}

\author{Michael Kesden$^{1}$ and Marc Kamionkowski$^2$}
\affiliation{$^1$Canadian Institute for Theoretical Astrophysics,
     University of Toronto, Toronto, ON M5S 3H8, Canada\\ 
     $^2$California Institute of Technology, Mail Code 130-33,
     Pasadena, CA 91125}

\date{\today}

\begin{abstract}
Satellite galaxies are tidally disrupted as they orbit the Milky Way.  If dark
matter (DM) experiences a stronger self-attraction than baryons, stars will
preferentially gain rather than lose energy during tidal disruption
leading to an enhancement in the trailing compared to the leading tidal stream.
The Sgr dwarf galaxy is seen to have roughly equal streams, challenging models
in which DM and baryons accelerate differently by more than 10\%.  Future
observations and a better understanding of DM distribution should allow
detection of equivalence violation at the percent level.
\end{abstract}

\pacs{98.65.Fz, 95.35.+d, 98.56.Wm, 98.10.+z}
\maketitle

Galileo showed at the Leaning Tower that objects of different
masses and materials fall the same way in a gravitational field.
This equivalence principle, later a cornerstone of Einstein's
general relativity, has been tested repeatedly through a variety
of experiments since Galileo's time and since the time of
Einstein (see, e.g., Ref.~\cite{cliffwill}).  These tests have
confirmed that the materials found
on Earth and in the Solar System all satisfy the equivalence
principle to a remarkable degree.  But what about
the dark matter that fills galactic halos and dominates the mass
density of the Universe?  Is the dark matter in the Milky
Way's halo accelerated the same as baryons in a gravitational
field?

The simplest and most favored candidates for dark matter (DM),
like weakly interacting massive particles
\cite{jkg,bergstrom,hooper} and axions
\cite{turner,raffelt,rosenberg}, do satisfy
the equivalence principle (EP).  However, there are a number of
reasons to test this assumption.  First of all, we still have no
empirical evidence for the existence of WIMPs or axions.  Some
have argued that a stronger self-gravity for dark matter is
required to clear dwarf galaxies from voids in the galaxy
distribution \cite{voids,string,STcosmo}.  Moreover, the recent
discovery that the cosmological expansion is accelerating
\cite{perlmutter,riess} suggests that there
may be more to gravity than general relativity---in particular,
the quintessence field may mediate an
additional long-range self-interaction between dark-matter
particles \cite{farrar,bias}.  There is thus considerable
motivation to scrutinize our cherished notions about the
equivalence principle.

Violations of the equivalence principle in the dark sector may
be modeled phenomenologically by attributing to dark-matter
particles $\psi$ a ``fifth force'' \cite{EPlett,frieman},
\begin{equation} \label{E:P1}
     V_{\phi}(r) = - \frac{g^2}{4\pi r} e^{-m_{\phi} r}.
\end{equation}
Here $g$ is a dimensionless coupling constant, and $m_{\phi}$ is
the mass of the scalar particle $\phi$ mediating the
interaction.  On scales $r \ll m_{\phi}^{-1}$, the potential of
Eq.~(\ref{E:P1}) leads to an inverse-square-law force between DM
particles of mass $m_{\psi}$ with a strength suppressed by a
factor $\beta^2$ compared to gravity, where $\beta \equiv
gm_{\rm Pl}/\sqrt{4\pi}m_{\psi}$.

Several cosmological consequences of such a DM force have
already been explored.  First of all, to clear dwarf galaxies
from voids, values $\beta \gtrsim 1$ and $m_{\phi}^{-1} \gtrsim
1$~Mpc are required \cite{nusser}.  An attractive
force for $r \ll m_{\phi}^{-1}$ would enhance structure formation on
these scales leading to a corresponding increase in the
density-perturbation power spectrum \cite{EPlong}, an
effect, though, that can be mimicked by a blue tilt in the
power spectrum.  An EP-violating coupling
between DM and quintessence could also induce a scale-independent bias
between baryons and DM, though this effect is model dependent
\cite{bias}.  Refs.~\cite{EPlett,EPlong} had
noted that a DM force would strip a baryonic core from its
dark halo, and applied this to typical galaxies in the Coma cluster
to set a limit $\beta < 2.2$.  Clusters might also test an attractive DM
force, as baryons would be preferentially lost compared to the
more tightly bound DM during the mergers leading to their
formation \cite{clusters}.  This test is complicated, however, by
gas physics which is expected to reduce the cluster baryon-to-DM
mass ratio below the cosmological value, even in the absence of
a DM force.

In this {\sl Letter}, we consider the effects of a DM force on
{\it galactic} scales.  We propose here that
tidal streams produced by the disruption of a DM-dominated
satellite galaxy orbiting in the halo of a much larger host
galaxy provide a powerful probe of an EP-violating DM force.
The reasoning follows by comparing the satellite's orbital
energy $E_{\rm orb}$, the energy $E_{\rm tid}$ imparted during
tidal disruption, and the self-binding energy $E_{\rm bin}$ of
the satellite \cite{TSdyn},
\begin{eqnarray} \label{E:3E}
     E_{\rm orb} &=& \frac{GM_R}{R} \, , \\
     E_{\rm tid} &=& r_{\rm tid} \frac{d\Phi_{\rm host}}{dR} =
     \left( \frac{m_{\rm sat}}{M_R} \right)^{1/3} E_{\rm orb} \,
     , \\ 
     E_{\rm bin} &=& \frac{Gm_{\rm sat}}{r_{\rm sat}} = \left(
     \frac{m_{\rm sat}}{M_R} \right)^{2/3} E_{\rm orb}.
\end{eqnarray}
Here the host galaxy has a potential $\Phi_{\rm host}(R)$ and mass
$M_R$ within the satellite's orbit of radius $R$, and the satellite has
a mass $m_{\rm sat}$ and radius $r_{\rm sat}$ which fill its tidal
radius $r_{\rm tid}$.
When the satellite is much less massive than the host galaxy,
$m_{\rm sat}/M_R \ll 1$, a distinct hierarchy,
\begin{equation} \label{E:hier}
     E_{\rm orb} \gg E_{\rm tid} \gg E_{\rm bin},
\end{equation}
exists in these three energy scales, implying that the disrupted
stars and satellite will trace similar orbits in the host
galaxy's potential regardless of the details of tidal disruption
or the satellite's internal structure.  The disrupted stars will
act like purely baryonic test particles, while the satellite
itself behaves largely like a DM test particle, if it is
DM dominated.

Fortunately, the Sagittarius (Sgr) dwarf galaxy, the Milky Way's
closest satellite at a Galactocentric distance of only 16 kpc,
is nearly ideal for our purposes.  The Sgr dwarf has extended
leading and trailing tidal streams observed by the Two-Micron All Sky
Survey (2MASS) \cite{2MassObs} and the Sloan Digital Sky Survey (SDSS)
\cite{SDSS}.  Using a sample of over 1,000 M-giant
stars with a known color-magnitude relation, the 2MASS
collaboration have measured not just surface brightnesses along
the streams, but distances and spectroscopic velocities as well
\cite{2MassVel}.  Comparing these observations to simulations
has led to estimates of the mass of the Sgr dwarf of $M_{\rm
Sgr} = (2-5) \times 10^8 M_{\odot}$, mass-to-light ratio $M_{\rm
Sgr}/L_{\rm Sgr} = 14-36\, M_\odot/L_\odot$, and Sgr orbit with
pericenter 10--19 kpc, apocenter 56--59 kpc, and period 0.85--0.87 Gyr
\cite{2MassTT}.  The large mass-to-light ratio suggests that the
Sgr dwarf is indeed DM dominated and therefore a suitable place
to search for DM forces.

To study more carefully the effects of EP violation on tidal disruption,
we performed our own simulations of the tidal disruption of a satellite
with a mass $(5 \times 10^8 M_{\odot})$, mass-to-light ratio
$(40 M_{\odot}/L_{\odot})$, and orbit (pericenter 14 kpc, apocenter 59
kpc) similar to that of the Sgr dwarf. We could not compare our
simulations directly with those of Ref.~\cite{2MassTT}, as we performed
$N$-body simulations of a Navarro-Frenk-White (NFW) profile for our
Milky Way halos, and they used a static logarithmic potential.
An active halo allows for dynamical friction over the course of
the simulation and possible backreaction on the halo due to the
DM force.  While we did not attempt to reproduce the detailed
features of the Sgr tidal streams, our simulations are
sufficient to demonstrate that even a small DM force could have
significant observational consequences.  The initial conditions
for our simulations were produced using GALACTICS
\cite{WidDub}, which makes use of phase-space distribution
functions (DFs) that are analytic in the orbital energy and
angular momentum.  By Jeans' theorem, these
DFs are equilibrium solutions to the collisionless Boltzmann
equations \cite{KuDub}, and they can be combined to produce
realistic and stable models of the composite Milky Way
bulge-disk-halo system \cite{WidDub}.  We used the two Milky Way
models of Ref.~\cite{WidDub} that best fit observational constraints,
including the Galactic rotation curve and local velocity
ellipsoid.  The simulations were evolved using a modified
version of the $N$-body code GADGET-2 \cite{Gadget2}.  A more
detailed description of our simulations are provided in
Ref.~\cite{KesKam2}.

\begin{figure}[t!]
\scalebox{0.40}{\includegraphics{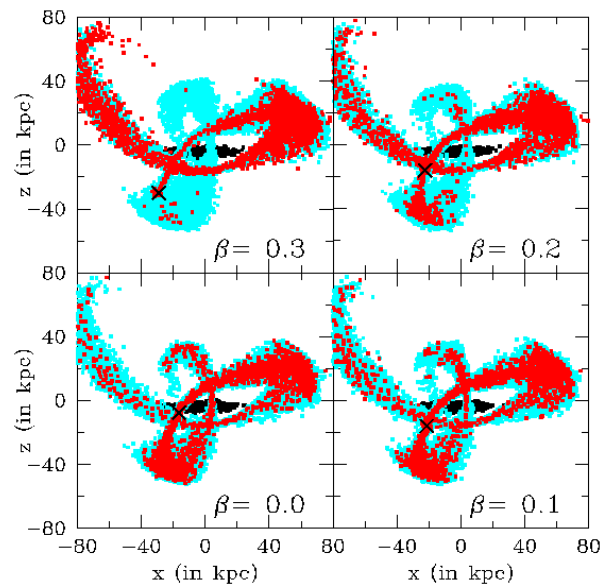}}
\caption{Simulations of the tidal disruption of a satellite
     galaxy in the presence of a dark-matter force. The charge-to-mass
     ratio $\beta$ increases from 0.0 in increments of 0.1 going
     counterclockwise from the bottom left.  The Galactic disk
     is in black.  Sgr stars are shown in
     red (dark grey) while the Sgr dark matter is blue (light grey).
     The tidal streams are projected onto the orbital plane.  Orbits
     are counterclockwise; the upper left figure shows that for
     $\beta = 0.3$ (a dark-matter force 9\% the strength of gravity)
     stars are almost absent from the leading stream (at 12 o'clock
     with respect to the Galactic center). X's denote the location of the
     bound Sgr core.}
\label{F:XZ}
\end{figure}

Four simulations of tidal disruption are depicted in
Fig.~\ref{F:XZ}, with DM forces given by Eq.~(\ref{E:P1}) with
different values of the charge-to-mass ratio $\beta$.  The
scalar field is assumed massless ($m_{\phi} = 0$), so the DM force is
a true inverse square law.  The ratio $\beta$ increases
from 0.0 at bottom left to 0.3 at top left as one proceeds
counterclockwise.  The simulations begin with the satellite at
apocenter 59 kpc from the Galactic center and last for 2.4 Gyr
(almost three full orbits).  The tangential velocities are
adjusted so that all orbits are projected to have a pericenter
of 14 kpc.  The orbits are counterclockwise in the $x$-$z$ plane, so
that the edge of the leading stream appears at 12 o'clock with
respect to the Galactic center in Fig.~\ref{F:XZ}, while the edge
of the trailing stream is at about 10 o'clock.  The Sgr dwarf is
modelled with a truncated NFW profile for both stars and DM, in
keeping with the simulations of Ref.~\cite{2MassTT}, where it
was concluded that observations could not yet
determine distinct profiles for the two components.  Thus, the
stars shown in red (dark grey) in the bottom left panel are
simply a downsampling of the DM distribution illustrated in blue
(light grey).

As the DM force increases in strength, the leading stream is
systematically depleted of stars, while the trailing stream is
correspondingly enhanced.  By the time $\beta$ reaches 0.3 in
the top left panel, the leading stream is virtually devoid of
stars.  The primary reason for this effect is that in the presence
of an attractive DM force, the center of mass of the satellite's
stars is displaced outwards with respect to that of its DM.  The
bound stars lie at the bottom of the satellite's gravitational
potential well and are therefore forced to orbit the Galactic
center at the same speed as the DM.  However, they do not have
the attractive pull of the DM force from the Milky Way's halo to
supplement gravity in providing the required centripetal force.
The stars are therefore displaced outwards so that the inward
gravitational pull of the satellite's DM can provide this
additional centripetal force.  From this outer position, stars
are more likely to be tidally disrupted from the far side of the
satellite than the side closest to the Galactic center.  Stars
disrupted from the far side gain energy and are boosted onto
higher orbits in the Milky Way's potential well where their
angular velocity is slower than that of the satellite.  They
therefore trail behind the satellite and develop into a trailing
tidal stream.  A repulsive DM force will induce an opposite
effect, displacing the stars towards the Galactic center and
preferentially creating a leading rather than a trailing tidal
stream.

\begin{figure}[t!]
\scalebox{0.40}{\includegraphics{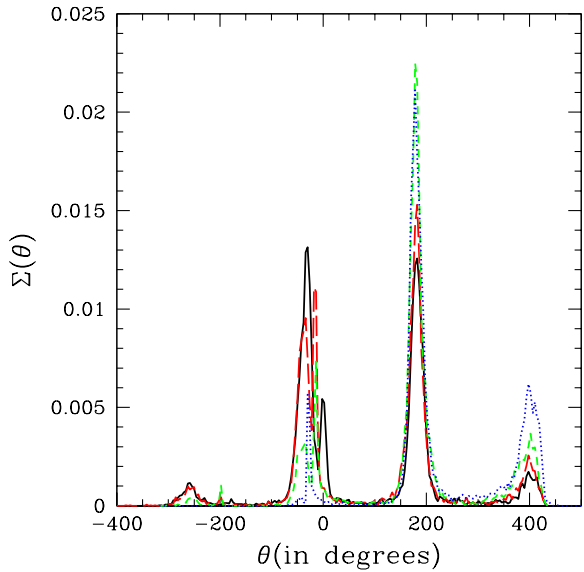}}
\caption{Surface density of stars as a function of angular
     distance $\theta$ along the tidal stream.  The satellite
     core is located at $0^{\circ}$, while the trailing and
     leading streams are at positive and negative $\theta$
     respectively.  As the tidal streams wrap around the Galaxy
     more than once, $\theta$ extends beyond $\pm 180^{\circ}$.
     The four curves correspond to the four panels of
     Fig.~\ref{F:XZ}, with black (solid), red (long-dashed),
     green (short-dashed), and blue (dotted) curves belonging to
     the $\beta = 0.0$ through 0.3 simulations.}
\label{F:dNdt} 
\end{figure}

\begin{figure}[t!]
\scalebox{0.40}{\includegraphics{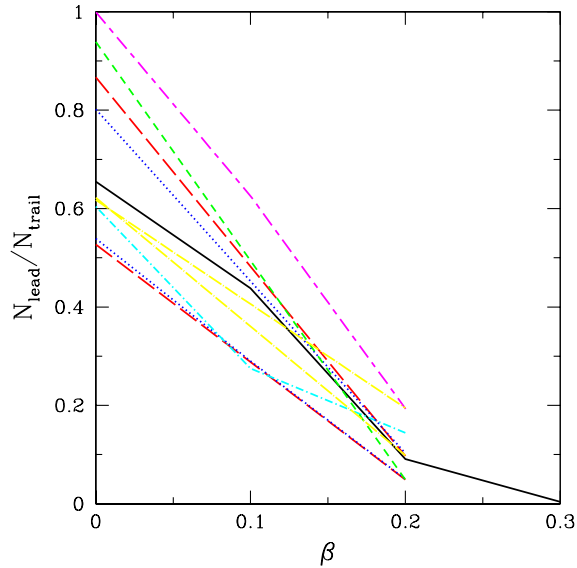}}
\caption{The ratio of leading to trailing stars as a function of
charge-to-mass ratio $\beta$ for different models of the host-satellite
system.  The black (solid) curve is our default best-fit model. The
magenta (long-short dashed) curve doubles the initial mass of the
satellite.   The two red (long-dashed) curves have rotating satellites:
the top is prograde and the bottom is retrograde.  The blue (dotted)
curves have satellites with different orbits: the top curve has a more
circular orbit, while the bottom curve has a planar orbit rather than
the polar orbit of Sgr.  The cyan (dot-short dashed) curve uses a Milky
Way model with lighter halo and heavier disk.  The green (short-dashed)
curve has a satellite where 25\% of the most bound particles represent
stars.  The yellow (dot-long dashed) curves have satellites with lower
$M/L$ ratios (higher stellar mass fractions): the top curve has
$M/L = 4.5$ while the bottom curve has $M/L = 10$.} \label{F:ratio}
\end{figure}

This asymmetry in the leading compared to the trailing tidal
streams is a distictive signature of a DM force that can be
observed in the stellar densities measured along the stream.
The normalized stellar densities of the four simulations
presented in Fig.~\ref{F:XZ} are shown in the four curves of
Fig.~\ref{F:dNdt}.  Orbits in our composite Milky Way model do
not close, and the four peaks in the stellar density distribution
correspond to the four apocenter passages appearing at 2, 7, 10,
and 12 o'clock in Fig.~\ref{F:XZ}.  Orbital velocities are
minimized at apocenter so stars tend to accumulate there.  The ratio
of the number of stars near the apocenters furthest along the leading
and trailing tidal streams thus provides a convenient measure of the
asymmetry between the streams.  The ratio of the number of stars in the
leading segment stretching from $-300^{\circ}$ to $-200^{\circ}$
as compared to the trailing segment from $350^{\circ}$ to
$450^{\circ}$ drops from 0.66 in the absence of a DM force down
to 0.44, 0.091, and 0.0042 for $\beta = 0.1$, 0.2, and 0.3 as indicated
by the solid black curve in Fig.~\ref{F:ratio}.
The SDSS has observed hundreds of stars per square degree along the
Sgr tidal stream \cite{SDSS}.  As the Sgr dwarf is observed to have an
extensive leading stream, we conclude that a DM force as weak as 9\% the
strength of gravity is likely to be observationally unacceptable, in which case
the proposal \cite{string,STcosmo,nusser} that EP-violating dark
matter clears dwarf galaxies from voids would be ruled out.

Our simulations suggest that current observations can already place
impressive constraints on a DM-force, but several concerns remain before
we can confront our simulations with data.  If the stream wraps around
the Galaxy more than once, we must be able to distinguish true leading
stars from trailing stars that have almost been lapped by the
satellite.  We have been able to do this surprisingly well in
simulations using only the radial velocities, distances, and positions
along the stream.  As 2MASS has collected this data, identifying leading
and trailing stars should already be feasible and can ceratainly be
accomplished by a future high-precision astrometry experiment like the
Space Interferometry Mission (SIM) or Gaia.  More troublesome is whether some
other change in our Sgr or Milky Way models could produce the same asymmetric
tidal tails that we are claiming as a signature of a DM force.
Future investigation of this concern is certainly needed, and we have
made a first attempt at this in Ref.~\cite{KesKam2}, the results of
which are summarized in Fig.~\ref{F:ratio}.  This signature is seen to
be robust to changes in the Milky Way model and the mass, orbit, and
phase-space distribution of the Sgr dwarf.  In the absence of a DM
force, the leading-to-trailing ratio exceeds 0.5 for all our models,
while for $\beta > 0.2$ the ratio is always below 0.2.  The detailed
morphology of the stream also allows us to anticipate when the
ratio should be high or low, making our test more sensitive than
a single number.  Our simulations suggest that it may
be possible to detect a DM force a few percent the strength of
gravity.  We may not be able to drop DM off the Leaning Tower of
Pisa, but the Sgr tidal streams may be the next best thing.

\begin{acknowledgments}
We wish to thank John Dubinski and Larry Widrow for assistance
with the use of GALACTICS and Pat McDonanld, Neal Dalal,
Christoph Pfrommer, and Jonathan Sievers for useful
conversations.  All computations were performed on CITA's
McKenzie cluster \cite{McKenzie}, which was funded by the Canada
Foundation for Innovation and the Ontario Innovation Trust.
Kesden acknowledges support from the NASA
Graduate Research Program, and NSERC of Canada.  Kamionkowski
acknowledges support from the DoE DE-FG03-92-ER40701, NASA
NNG05GF69G, and the Gordon and Betty Moore Foundation.
\end{acknowledgments}

\bibliography{lett}

\end{document}